\documentclass[12pt]{article}
\begin{document}

\title{Entanglement with phase decoherence}
\author{Jing-Bo Xu$^{1,2}$ and Shang-Bin Li$^{2}$  \\
$^{1}$Chinese Center of Advanced Science and Technology (World Labor-\\
atory), P.O.Box 8730, Beijing, People's Republic of China;\\
$^{2}$Zhejiang Institute of Modern Physics and Department of Physics,\\
Zhejiang University, Hangzhou 310027, People's Republic of
China\thanks{Mailing address}}
\date{}
\maketitle

\begin{abstract}
{\normalsize The system of an atom couples to two distinct optical
cavities with phase decoherence is studied by making use of a
dynamic algebraic method. We adopt the concurrence to characterize
the entanglement between atom and cavities or between two optical
cavities in the presence of the phase decoherence. It is found
that the entanglement between atom and cavities can be controlled
by adjusting the detuning parameter. Finally, we show that even if
the atom is initially prepared in a maximally mixed state, it can
also entangle the two mode cavity fields.\\

}
\end{abstract}

\newpage
\section * {I. INTRODUCTION}
\hspace*{8mm}Quantum entanglement was first introduced by
Einstein, Podolsky and Rosen in their famous paper in 1935 [1]. It
has been recently recognized that entanglement can be used as an
important resource for quantum information processing [2].
Entanglement can exhibit the nature of a nonlocal correlation
between quantum systems that have no classical interpretation.
However, real quantum systems will unavoidably be influenced by
surrounding environments. The interaction between the environment
and quantum systems of interest can lead to decoherence. It is
therefore of great importance to prevent or minimize the influence
of environmental noise in the practical realization of quantum
information processing. In order to prevent the effect of
decoherence, several approaches have been proposed such as quantum
error-correcting approach [3] or quantum error-avoiding approach
[4,5].\\
\hspace*{8mm}Instead of attempting to shield the system from the
environmental noise, Plenio and Huelge [6] use white noise to play
a constructive role and generate the controllable entanglement by
incoherent sources. Similar work on this aspect has also been
considered by other authors [7]. In this paper, we investigate an
atom couples to two distinct optical cavities with the phase
decoherence and show how the entanglement between atom and
cavities or between two optical cavities can be generated in the
presence of the phase decoherence. In section \textrm{II}, we
study the system with phase decoherence by making use of the
dynamic algebraic method [8] and find the exact solution of the
master equation for the system. The exact solution is then used to
discuss the influence of the phase decoherence on the probability
of occupation in ground state. In section \textrm{III}, we use the
concurrence to characterize the entanglement between atom and
cavities or between two optical cavities by means of the exact
solution for the system. It is shown that the entanglement between
atom and cavities can be controlled by adjusting the detuning
parameter. Finally, we show that even if the atom is initially
prepared in a maximally mixed state, it can also entangle the two
mode cavity fields. A conclusion is given in Section \textrm{IV}.\\

\section * {II. SOLUTION OF AN ATOM COUPLES TO TWO DISTINCT OPTICAL CAVITIES WITH PHASE DECOHERENCE}
\hspace*{8mm}We consider the situation that an atomic system is
surrounded by two distinct optical cavities initially prepared in
the vacuum state. The Hamiltonian for the system can be described
by [6],
$$
H=\omega_aa^{\dagger}a+\omega_bb^{\dagger}b+\frac{\omega_0}{2}(|e\rangle\langle{e}|
-|g\rangle\langle{g}|)+g_a(a|e\rangle\langle{g}|+a^{\dagger}|g\rangle\langle{e}|)
+g_b(b|e\rangle\langle{g}|+b^{\dagger}|g\rangle\langle{e}|),
\eqno{(1)}
$$
where $|e\rangle$ and $|g\rangle$ are the excited state and the
ground state of the two-level atom, $\omega_0$ is atomic
transition frequency, $g_{a(b)}$ is the coupling constant of the
atom to cavity modes $a(b)$, and $a$ ($a^{\dagger}$), $b$
($b^{\dagger}$) are the annihilation (creation) operators of $a$
mode of frequency $\omega_a$ and $b$ mode of frequency $\omega_b$,
respectively. In Ref.[6], Plenio and Huelge use white noise as the
actual driving force of the system and study numerically the
entanglement between two optical cavities for the system in the
resonant case. Here, we investigate analytically the entanglement
between atom and cavities or between two optical cavities with
phase decoherence by making use of the dynamical algebraical
method. To reduce the complexity, we consider the case of
$\omega_a=\omega_b=\omega$. It is easy to verify that there exists
two constants of motion in Hamiltonian (1),
$$
K_1=\frac{1}{g^{2}}(g^2_a{a}^{\dagger}a+g^2_b{b}^{\dagger}b)+\frac{g_ag_b}{g^2}(a^{\dagger}b
+ab^{\dagger})+\frac{1+|e\rangle\langle{e}|-|g\rangle\langle{g}|}{2},
$$
$$
K_2=\frac{1}{g^{2}}(g^2_a{b}^{\dagger}b+g^2_b{a}^{\dagger}a)-\frac{g_ag_b}{g^2}(a^{\dagger}b
+ab^{\dagger}), \eqno{(2)}
$$
where $g=\sqrt{g^2_a+g^2_b}$. It is easily proved that the
operator $K_1$ and $K_2$ commute with Hamiltonian (1). We then
introduce the operators as follows
$$
S_+=\frac{(g_a{a}+g_b{b})|e\rangle\langle{g}|}{g\sqrt{K_1}},
~~~S_-=\frac{(g_a{a^{\dagger}}+g_b{b^{\dagger}})|g\rangle\langle{e}|}{g\sqrt{K_1}},
$$
$$
S_0=\frac{1}{2}(|e\rangle\langle{e}|-|g\rangle\langle{g}|).
\eqno{(3)}
$$
It can be shown that the operators $S_i$ ($i=0,\pm$) satisfy the
following commutation relations
$$
[S_0,S_{\pm}]=\pm{S}_{\pm},~~~[S_+,S_-]=2S_0,
\eqno{(4)}
$$
where $S_0$ and $S_{\pm}$ are the generators of the SU(2) algebra
[9]. In terms of the SU(2) generators, we can rewrite Hamiltonian
(1) as
$$
H=\omega(K_1+K_2-\frac{1}{2})+\Delta{S}_0+g\sqrt{K_1}(S_{+}+S_{-}),
\eqno{(5)}
$$
where $\Delta=\omega_0-\omega$ denotes detuning. In this paper, we
consider the pure phase decoherence mechanism only. In this
situation, the master equation governing the time evolution for
the system under the Markovian approximation is given by [10]
$$
\frac{d\rho}{dt}=-i[H,\rho]-\frac{\gamma}{2}[H,[H,\rho]],
\eqno{(6)}
$$
where $\gamma$ is the phase decoherence rate. Noted that the
equation with the similar form has been proposed to describing the
intrinsic decoherence [11]. The formal solution of the master
equation (6) can be expressed as follows [8],
$$
\rho(t)=\sum^{\infty}_{k=0}\frac{(\gamma{t})^{k}}{k!}M^{k}(t)\rho(0)M^{\dagger{k}}(t),
\eqno{(7)}
$$
where $\rho(0)$ is the density operators of the initial atom-field
system and $M^{k}(t)$ is defined by
$$
M^k(t)=H^k\exp(-iHt)\exp(-\frac{\gamma{t}}{2}H^2).
\eqno{(8)}
$$
By means of the SU(2) dynamical algebraic structure, we obtain the
explicit expression for the operator $M^k$
$$
M^k(t)=\frac{1}{2}[f_{+}(K_1,K_2)]^{k}\exp[-if_{+}(K_1,K_2)t]
\exp[-\frac{\gamma{t}}{2}[f_{+}(K_1,K_2)]^2]
$$
$$
~~~+\frac{1}{2}[f_{-}(K_1,K_2)]^k\exp[-if_{-}(K_1,K_2)t]
\exp[-\frac{\gamma{t}}{2}[f_{-}(K_1,K_2)]^2]
$$
$$
~~~+\frac{1}{2}[\frac{\Delta}{\Omega(K_1)}(|e\rangle\langle{e}|-|g\rangle\langle{g}|)
+\frac{2H_{int}}{\Omega(K_1)}]\{[f_{+}(K_1,K_2)]^{k}\exp[-if_{+}(K_1,K_2)t]
\exp[-\frac{\gamma{t}}{2}[f_{+}(K_1,K_2)]^2]
$$
$$
~~~-[f_{-}(K_1,K_2)]^{k}\exp[-if_{-}(K_1,K_2)t]\exp[-\frac{\gamma{t}}{2}[f_{-}(K_1,K_2)]^2]\},
\eqno{(9)}
$$
where
$$
f_{\pm}(K_1,K_2)=\omega(K_1+K_2-\frac{1}{2})\pm\frac{1}{2}\Omega(K_1),
~~~\Omega(K_1)=(\Delta^2+4g^2K_1)^{1/2},
$$
$$
H_{int}=g_a(a|e\rangle\langle{g}|+a^{\dagger}|g\rangle\langle{e}|)
+g_b(b|e\rangle\langle{g}|+b^{\dagger}|g\rangle\langle{e}|).
\eqno{(10)}
$$
We assume that the cavity fields are prepared initially in vacuum
state $|00\rangle$, and the atom is prepared in the excited state
$|e\rangle$. The time evolution of $\rho(t)$ can be written as,
$$
\rho(t)=\frac{1}{2}[1+\frac{\Delta^2}{\Omega^2}+(1-\frac{\Delta^2}{\Omega^2})\cos{\Omega
t}\exp(-\frac{\gamma{t}}{2}\Omega^2)]|00\rangle\langle00|\otimes|e\rangle\langle{e}|
$$
$$
+\frac{g}{\Omega}\{\frac{\Delta}{\Omega}[1-\cos{\Omega{t}}\exp(-\frac{\gamma{t}}{2}\Omega^2)]
+i\sin\Omega{t}\exp(-\frac{\gamma{t}}{2}\Omega^2)\}
|00\rangle\langle{\varphi}|\otimes|e\rangle\langle{g}|
$$
$$
+\frac{g}{\Omega}\{\frac{\Delta}{\Omega}[1-\cos{\Omega{t}}\exp(-\frac{\gamma{t}}{2}\Omega^2)]
-i\sin\Omega{t}\exp(-\frac{\gamma{t}}{2}\Omega^2)\}
|\varphi\rangle\langle{00}|\otimes|g\rangle\langle{e}|,
$$
$$
~~~+\frac{2g^2}{\Omega^2}[1-\cos{\Omega t}
\exp(-\frac{\gamma{t}}{2}\Omega^2)]|\varphi\rangle\langle{\varphi}|\otimes|g\rangle\langle{g}|
\eqno{(11)}
$$
where
$$
|\varphi\rangle=\frac{1}{g}(g_a|10\rangle+g_b|01\rangle),
~~~\Omega=(\Delta^2+4g^2)^{1/2}. \eqno{(12)}
$$
The $|\varphi\rangle$ in Eq.(12) is a single-photon entangled
state. Recently, much attention has been paid to investigate the
preparation of the single-photon maximally entangled state [12].
It is noted that when the two coupling coefficients $g_a=g_b$, the
state $|\varphi\rangle$ is nothing but a single-photon maximally
entangled state. We then show that if a projective measurement on
the atom in the $\{|e\rangle,|g\rangle\}$ basis is made, the atom
will be projected on the ground state $|g\rangle$ with the
probability $P_g$ in the case of $\Delta=0$,
$$
P_g=\frac{1}{2}[1-\cos(2gt)\exp(-2\gamma g^2t)]. \eqno{(13)}
$$
If the measurement result is $|g\rangle$, the two distinct cavity
fields are in the single-photon maximally entangled state
$\frac{\sqrt{2}}{2}(|10\rangle+|01\rangle)$. In Fig.1, we plot the
probability $P_g$ as the function of time $t$ for different values
of phase decoherence rate $\gamma$. It is shown that if the
decoherence rate $\gamma$ is zero, the two distinct cavity fields
are in the maximally entangled single-photon state at the time
$t=\frac{\pi}{2g}$ with unit probability.\\

\section * {III. THE ENTANGLEMENT BETWEEN ATOM AND CAVITIES OR TWO OPTICAL CAVITIES}
\hspace*{8mm}In order to quantify the degree of entanglement,
several measures [13] of entanglement have been introduced for
both pure and mixed quantum states. In this section, we adopt the
concurrence to calculate the entanglement between atom and
cavities or between two optical cavities with the phase
decoherence. The concurrence related to the density operator
$\rho$ of a mixed state is defined by [14]
$$
C(\rho)=\max\{\lambda_1-\lambda_2-\lambda_3-\lambda_4,0\},
\eqno{(14)}
$$
where the $\lambda_i$ ($i=1,2,3,4$) are the square roots of the
eigenvalues in decreasing order of magnitude of the "spin-flipped"
density operator $R$
$$
R=\rho(\sigma_y\otimes\sigma_y)\rho^{\ast}(\sigma_y\otimes\sigma_y),
\eqno{(15)}
$$
where the asterisk indicates complex conjugation. The concurrence
varies from $C=0$ for an unentangled state to $C=1$ for a
maximally entangled state.\\
\hspace*{8mm}We first investigate the quantum correlation between
the atom and cavity modes. If we deal with the two cavity modes as
system B, and the atom as system A, then $\rho(t)$ in Eq.(11) can
be thought of as the density operator of a two-qubit mixed state.
In the basis $|11\rangle_s\equiv|00\rangle\otimes|e\rangle$,
$|10\rangle_s\equiv|00\rangle\otimes|g\rangle$,
$|01\rangle_s\equiv|\varphi\rangle\otimes|e\rangle$,
$|00\rangle_s\equiv|\varphi\rangle\otimes|g\rangle$, the explicit
expression of the concurrence $C_{AB}$ describing the entanglement
between the system A and system B can be found to be,
$$
C_{AB}=\frac{2g}{\Omega}\{\frac{\Delta^2}{\Omega^2}[1-\cos{\Omega
t} \exp(-\frac{\gamma{t}}{2}\Omega^2)]^2+\sin^2{\Omega
t}\exp(-\gamma{t}\Omega^2)\}^{1/2}. \eqno{(16)}
$$
From Eq.(16), we can see that the detuning $\Delta$ plays a key
role in the quantum correlation between the atom and cavity modes.
If the decoherence rate $\gamma$ is not equal to zero, the
concurrence $C_{AB}$ remains in the value $2g|\Delta|/\Omega^2$ in
the limit $t\rightarrow\infty$. In the strong coupling case, i.e.,
$g_a,g_b\gg\Delta$, the concurrence $C_{AB}$ of the stationary
state $\rho(\infty)$ is approximately $|\Delta|/(2g)$. On the
other hand, in the large detuning limit, the concurrence $C_{AB}$
of the stationary state is approximately $2g/|\Delta|$. In Fig.2,
the concurrence $C_{AB}$ is plotted as a function of the time $t$
and decoherence rate $\gamma$.  We show the concurrence $C_{AB}$
as a function of the detuning parameter $\Delta$ and the
decoherence rate $\gamma$ at a fixed time in Fig.3. In the limit
$t\rightarrow\infty$, concurrence $C_{AB}$ is plotted as a
function of the detuning parameter $\Delta$ in Fig.4. From Fig.4,
we can see that $C_{AB}$ increases with the detuning $\Delta$
parameter. This means that the entanglement between the atom and
the cavity fields can be controlled by adjusting the detuning
parameter. Now, we turn our discussion to the resonant case, i.e.
$\Delta=0$. In this case,
$C_{AB}=|\sin(2gt)|\exp(-2g^2\gamma{t})$.\\
\hspace*{8mm}In Ref.[15], it has been proved that for any pure
states of three qubits 1, 2 and 3, the entanglement is distributed
following the inequality for the squared concurrence
$$
C^2_{12}+C^2_{13}\leq{C}^{2}_{1(23)},
\eqno{(17)}
$$
where $C_{1,(23)}$ is the single-qubit concurrence defined as the
concurrence between the qubit 1 and the rest of qubits (2,3). For
any mixed states of three qubits 1, 2, and 3, there is analogous
inequality for the squared concurrence as follows
$$
C^2_{12}+C^2_{13}\leq\langle{C}^{2}\rangle^{min}_{1(23)},
\eqno{(18)}
$$
where $\langle{C}^{2}\rangle^{min}_{1(23)}$ is the minimum of
average over all possible pure state decomposition of the three
qubits mixed state [15].\\
\hspace*{8mm}In the present paper, we may expect that the pair
entanglement between the atom and the $a$($b$) mode cavity field
is determined by the coupling coefficient $g_a$($g_b$). It is easy
to prove that there exist the simple relations,
$$
C^2_{AB}=C^2_{a}+C^2_{b},
\eqno{(19)}
$$
and $C_a/C_b=g_a/g_b$, where $C_{a}(C_{b})$ is the concurrence
describing entanglement between the atom and the $a$($b$) mode
cavity field. Thus, our result is in agreement with that obtained
in Ref.[15].\\
\hspace*{8mm}Next, we investigate the entanglement between light
fields of two distinct cavities. By tracing out the degree of
freedom of the atom in density matrix $\rho(t)$ in Eq.(11), we
obtain the reduced density matrix $\rho_{B}(t)$ describing the two
light fields as follows,
$$
\rho_{B}(t)=\frac{1}{2}[1+\frac{\Delta^2}{\Omega^2}+(1
-\frac{\Delta^2}{\Omega^2})\cos{\Omega t}
\exp(-\frac{\gamma{t}}{2} \Omega^2)]|00\rangle\langle00|,
$$
$$
~~~+\frac{2g^2}{\Omega^2}[1-\cos{\Omega t}
\exp(-\frac{\gamma{t}}{2}\Omega^2)]|\varphi\rangle\langle\varphi|.
\eqno{(20)}
$$
Then, the concurrence $C_{B}$ characterizing the entanglement of
two light fields can be derived as
$$
C_{B}=\frac{4|g_ag_b|}{\Omega^2}[1-\cos{\Omega t}
\exp(-\frac{\gamma{t}}{2}\Omega^2)]. \eqno{(21)}
$$
From Eq.(21), we can see that the concurrence $C_{B}$ is equal to
zero at time $t=2n\pi/\Omega,(n=0,1,2...)$ in the case of
$\gamma=0$. At these specific time, the two cavity modes have no
pair entanglement. However, in the case with $\gamma\neq0$, the
two cavity modes is always entangled for the time $t>0$. In Fig.5,
we plot the concurrence $C_{B}$ as the function of time $t$ and
damping rate $\gamma$. From Fig.5, we see that the entanglement
between the two distinct light fields increases with the phase
decoherence rate $\gamma$ within the time range
$2n\pi/\sqrt{\Delta^2+4g^2}\leq{t}<(2n+\frac{1}{2})\pi/\sqrt{\Delta^2+4g^2}$
or
$(2n+\frac{3}{2})\pi/\sqrt{\Delta^2+4g^2}<{t}\leq(2n+2)\pi/\sqrt{\Delta^2+4g^2}$
($n=0,1,2,...$). The concurrence $C_B$ is displayed as a function
of the phase decoherence rate $\gamma$ for three different values
of the detuning parameters at fixed time in Fig.6. The stationary
state entanglement of the two cavity modes measured by concurrence
is $4g_ag_b/(\Delta^2+4g^2)$. This means that the stationary state
entanglement achieves its maximal value $1/2$ in the resonant case
with $g_a=g_b$.\\
\hspace*{8mm}Finally, we discuss how much entanglement between the
two mode cavity fields can be achieved if the initial atom is
prepared in a thermal state and the cavity fields are prepared in
the vacuum states. We assume that the initial atom is in the state
$\rho_A(0)=\delta|g\rangle\langle{g}|+(1-\delta)|e\rangle\langle{e}|$,
where $0\leq\delta\leq1$, and the cavity fields are still in the
vacuum state $|00\rangle$. Our calculation shows that
$C^{\prime}_{AB}=(1-\delta)C_{AB}$ and
$C^{\prime}_{B}=(1-\delta)C_{B}$. This means that even if the
initial atom is prepared in a maximally mixed state
$\frac{1}{2}|g\rangle\langle{g}|+\frac{1}{2}|e\rangle\langle{e}|$,
it can still entangle the two mode cavity fields. In this case,
the concurrence $C^{\prime}_{B}$ equals $\frac{1}{4}$ in the
steady state for $\Delta=0$ and $g_a=g_b$.\\

\section * {IV. CONCLUSION}
\hspace*{8mm}In this paper, we investigate analytically the
entanglement between atom and cavities or between two optical
cavities with phase decoherence by making use of the dynamic
algebraic method. It is found that the entanglement between atom
and cavities can be controlled by adjusting the detuning
parameter. Finally, we show that even if the atom is initially in
a maximally mixed state, it can also entangle two mode cavity
fields initially prepared in vacuum state. The approach adopted
here can be employed to investigate the entanglement between two
optical cavities mediated by a two-level atom in those cases, in
which the two mode cavity fields are initially prepared in another
separable states.\\

\section * {ACNOWLEDGMENT}
This project was supported by the National Natural Science
Foundation of China (Project NO. 10174066).

\newpage
{\Large\bf Figure Caption}
\begin{description}
\item[FIG.1.]The probability $P_g$ as a function of the time $t$ for various values phase
damping rate : $\gamma=1$ (Solid), $\gamma=0$ (dash), $\gamma=0.01$ (dot) and
$\gamma=0.05$ (dash dot) with $g_a=g_b=1$ and $\Delta=0$.\\
\item[FIG.2.]The concurrence $C_{AB}$ as a function of the time $t$ and the phase
damping rate $\gamma$  for $g_a=g_b=1$ and $\Delta=5$.\\
\item[FIG.3]The concurrence $C_{AB}$ as a function of the detuning parameter $\Delta$ and
the phase damping rate $\gamma$  for $g_a=g_b=1$ and $t=10$.\\
\item[FIG.4]The concurrence $C_{AB}$ of the steady state as a function of the
detuning parameter $\Delta$ for $g_a=g_b=1$ and $\gamma=0.1$.\\
\item[FIG.5]The concurrence $C_{B}$ as a function of the time $t$ and the phase damping
rate $\gamma$ for $g_a=g_b=1$ and $\Delta=0$.\\
\item[FIG.6]The concurrence $C_{B}$ as a function of the phase damping rate $\gamma$
for various values of the detuning parameter: $\Delta=0$ (Solid), $\Delta=1$ (Dash) and
$\Delta=2$ (Dot) with $t=2$ and $g_a=g_b=1$.\\
\end{description}


\newpage
\begin{thebibliography}{99}
\bibitem{1}A. Einstein, B. Podolsky and N. Rosen, Phys. Rev. \textbf{47},
777(1935).
\bibitem{2}D. P. DiVincenzo, Science \textbf{270}, 255 (1995); L. K. Grover,
Phys. Rev. Lett. \textbf{79}(2), 325 (1997); J.I. Cirac and P.
Zoller, Nature \textbf{404}, 579 (2000); M. A. Nielsen and I. L.
Chuang, \textit{Quantum Computation and Quantum Information}
(Cambridge University Press, Cambridge, 2000).
\bibitem{3}P. W. Shor, Phys. Rev. A \textbf{52}, 2493(1995); A. M.
Steane, Phys. Rev. Lett. \textbf{77}, 793 (1996); E. Knill and R.
Laflamme, Phys. Rev. A \textbf{55}, 900 (1997).
\bibitem{4}L.-M. Duan and G.-C. Guo, Phys. Rev. Lett.
\textbf{79}, 1953 (1997); P. Zanardi and M. Rasetti, Phys. Rev.
Lett. \textbf{79}, 3306 (1997); D. A. Lidar, I. L. Chuang, and K.
B. Whaley, Phys. Rev. Lett. \textbf{81}, 2594 (1998).
\bibitem{5}H. M. Wiseman and G. J. Milburn, Phys. Rev. Lett. \textbf{70}, 548
(1993); D. Vitali
\bibitem{6}M. B. Plenio, S. F. Huelga, Phys. Rev. Lett. \textbf{88},
197901 (2002).
\bibitem{7}A. Beige, S. Bose, D. Braun, S.F. Huelga, P.L. Knight, M.B. Plenio,
and V. Vedral, J. Mod. Opt. \textbf{47}, 2583 (2000); M. S. Kim,
J. Lee, D. Ahn, P. L. Knight, Phys. Rev. A \textbf{65}, 040101(R)
(2002).
\bibitem{8}Jing-Bo Xu and Xu-Bo Zou, Phys. Rev. A \textbf{60}, 4743 (1999).
\bibitem{9}Y. Wu, Phys. Rev. A \textbf{54}, 4534 (1996); Jing-Bo Xu, Xu-Bo Zou
and Ji-Hua Yu, Eur. Phys. J. D \textbf{10}, 295 (2000); Y. Wu and
X. Yang, Phys. Rev. A \textbf{63}, 043816 (2001).
\bibitem{10}C. W. Gardiner, \textit{Quantum Noise} (Springer-Verlag,
Berlin, 1991); W. H. Louisell, \textit{Quantum Statistics
Properties of Radiation} (Wiley, New York, 1973).
\bibitem{11}G. M. Milburn, Phys. Rev. A \textbf{44}, 5401 (1991)
\bibitem{12}M. D. Lukin, A. Imamo\v{g}lu, Phys. Rev. Lett. \textbf{84}, 1419 (2000);
J. C. Howell, J. A. Yeazell, Phys. Rev. Lett. \textbf{85}, 198
(2000); A. Rauschenbeutel \textit{et} \textit{al}., Phys. Rev. A
\textbf{64}, 050301 (2001);
\bibitem{13}W. K. Wootters, Quantum Inf.
Comput. \textbf{1}, 27 (2001), and references therein.
\bibitem{14}W. K. Wootters, Phys. Rev. Lett. \textbf{80}, 2245
(1998).
\bibitem{15}V. Coffman, J. Kundu, W. Wootters, Phys. Rev. A
\textbf{61}, 052306 (2000).
\end{thebibliography}
\end{document}